\documentclass[aps,prd,eprint,numerical,showpacs,showkeys,10pt]{revtex4-2}

\usepackage{longtable}
\usepackage{amsmath}
\usepackage{amssymb}
\usepackage{amsfonts}
\usepackage{graphicx}
\usepackage{dcolumn}
\usepackage{bm}
\usepackage{hyperref}
\usepackage{color}
\hypersetup{colorlinks=false}

\begin{document}

\title[Quasibound states of Unruh's acoustic black hole]{Quasibound states of analytic black-hole configurations in three and four dimensions}

\date{\today}

\author{H. S. Vieira}
\email{horacio.santana.vieira@hotmail.com}
\email{horacio_vieira@ufla.br}
\email{horacio.santana-vieira@tat.uni-tuebingen.de}
\email{horacio.santana.vieira@ifsc.usp.br}
\affiliation{Department of Physics, Institute of Natural Sciences, Federal University of Lavras, 37200-000 Lavras, Brazil}
\affiliation{Theoretical Astrophysics, Institute for Astronomy and Astrophysics, University of T\"{u}bingen, 72076 T\"{u}bingen, Germany}
\affiliation{S\~{a}o Carlos Institute of Physics, University of S\~{a}o Paulo, 13560-970 S\~{a}o Carlos, Brazil}

\begin{abstract}

In this work we analyze the sound perturbation of Unruh's acoustic effective geometry in both (2+1) and (3+1) spacetime dimensions and present an exact analytical expression for the quasibound states of these idealized black-hole configurations by using a new approach recently developed, which uses the polynomial conditions of the hypergeometric functions. Our main goal is to discuss the effects of having an event horizon in such effective metrics. We also discuss the stability of the systems and present the radial eigenfunctions related to these quasibound state frequencies. These metrics assume just the form it has for a Schwarzschild black hole near the event horizon, and therefore may, in principle, shed some light into the underlying classical and quantum physics of astrophysical black holes through analog acoustic probes.

\end{abstract}



\preprint{Preprint submitted to EPJC}
%\preprint{AIP/123-QED}

\maketitle

%\begin{quotation}
%...
%\end{quotation}

%
%%%%%%%%%%%%%%%%%%%%%%%%%%%%%%%%%%%%%%%%%%%%%%%%%%%%%%%%%%%%%%%%% Introduction
%
\section{Introduction}\label{Introduction}

The theory of general relativity had predicted some astrophysical and cosmological phenomena, as for example the gravitational redshift and Mercury's perihelion \cite{d'Inverno:1998}, which had passed for experimental tests with high precision. Furthermore, another very interesting, important prediction is the existence of black holes. From a theoretical point of view, the black-hole physics has been a very productive field of research in the last century, including the studies of classical and quantum properties of these objects \cite{LivingRevRelativity.2.2,LivingRevRelativity.22.4}. Henceforth, these properties have been passing observational checks with very high accuracy, as for example the gravitational waves emitted by compact objects and their binary systems \cite{PhysRevLett.116.221101,AstrophysJLett.875.L1}.

From a classical point of view, even the light waves cannot escape from the black-hole (final) attraction \cite{MTW:1973}. However, from a quantum point of view, the black holes can emit radiation with a specific (Hawking) temperature \cite{CommunMathPhys.43.199}. These physical phenomena are intrinsically related to the existence of an (exterior) event horizon \cite{Hawking:1973}. Two other forms of radiation that play a crucial role in understanding the physics of black holes are the quasinormal modes (QNMs) and the quasibound states (QBSs). The QNM frequencies are the set of vibrational spectra associated with the (damped reverberation) ringdown of a black hole \cite{ClassQuantumGrav.26.163001,RevModPhys.83.793}, which can be calculated when purely ingoing (outgoing) boundary conditions are imposed at the exterior event horizon (infinity).

The QBSs are localized mode solutions in the black-hole potential wells \cite{PhysLett.52B.437,LettNuovoCim.15.257}. They are ingoing waves at the exterior event horizon, which means that the radial solution diverges by reaching a maximum value, and tend to zero far from the black hole at the spatial infinity \cite{PhysRevD.76.084001}, which means that the probability of finding any particle there is null. In this context, the continuous spectrum of the classical stable bound states is replaced by a discrete spectrum of resonances with the tunneling through the potential barrier giving the finite probability of the particle to be captured by the horizon. That is the gravitational analog of the hydrogen atom orbitals, which means a spectrum of normalizable bound states. There has been, strangely, little effort devoted to the study of the quasibound state spectrum, despite the fundamental importance for the electromagnetic analog. Indeed, it is clear that these states must exist and they can provide a quantum description of a test particle orbiting a black hole. The physics of this problem can be understood if we assume that the singularity at the center of a black hole acts as a current sink, and therefore all normalizable states must decay in time, which implies that we must search for eigenstates over the two-dimensional space of complex energies. The constructed states have a finite half-life, so they can be viewed as resonance states (or quasistationary states). Thus, the QBS frequencies are the set of resonance spectra associated with the dissipation of energy at the exterior event horizon, and hence they are inevitably complex, which can be expressed as $\omega_{n}=\mbox{Re}[\omega]+i\mbox{Im}[\omega]$, where $n$ is the overtone number, $\mbox{Re}[\omega]$ is the real part (oscillation frequency), and $\mbox{Im}[\omega]$ is the imaginary part (decay or growth rate).

Obviously the observational measurements of some physical phenomena such as superradiance and Hawking radiation are almost impossible to make at the level of terrestrial laboratories with the current ground-based detectors, since the Hawking temperature is 7 orders of magnitude smaller than the temperature of the cosmic microwave background (CMB), for a Schwarzschild-like black hole with a mass equivalent to the solar mass. On the other hand, for the QNM spectrum, these difficulties are related to the process of extracting quasinormal information from gravitational waves. Finally, for the QBS spectrum, since it does not contain outgoing waves at spatial infinity, it would not leave any imprint in gravitational waves, and hence it could only be detected by analyzing the gravitational potential in the vicinity of a (relaxing) collapsing star. However, there have been several works proposing new (summation) methods to help detect these kinds of weak signals \cite{PhysRevD.93.044048,PhysRevD.98.024052}. Therefore, this fact widely opens the interest, and research as well, in analog models of gravity that mimic some properties of black-hole spacetimes.

Analog models of gravity were proposed by Unruh \cite{PhysRevLett.46.1351} as controlled tabletop laboratory experiments aiming to test some process that occurs in the interplay between general relativity and quantum physics. In this framework, the equation of motion describing the propagation of sound modes (the phonons) on a (supersonic) fluid flow can be identically written as the Klein-Gordon equation for a massless scalar field minimally coupled to an effective geometry containing a sonic event horizon; this kind of background is called an acoustic black hole. The most promising experiments with analog models of gravity were performed with fluids \cite{PhysRevLett.106.021302,PhysRevD.91.124018,PhysRevLett.117.121301,PhyRevLett.117.271101,NatPhys.13.833,PhysRevLett.121.061101,PhysRevLett.124.141101,PhysRevLett.126.041105,ClassQuantumGrav.36.194002,PhysRevLett.125.011301,PhysRevD.102.084041}, Bose-Einstein condensates \cite{PhysRevA.63.023611,NatPhys.10.864,Nature.569.688,PhysRevLett.125.213603,PhysRevResearch.4.023099}, and optical systems \cite{Optica.5.1099,PhysRevLett.122.010404}, among others (see Ref. \cite{PhilTransRSocA.378.20190239} and references therein).

In the present work, we analytically compute the QBSs (or resonant frequencies) in the (3+1)-dimensional Unruh's acoustic black hole by using the Vieira-Bezerra-Kokkotas (VBK) approach \cite{AnnPhys.373.28,PhysRevD.104.024035}, which suggests imposing the polynomial condition of Gauss's hypergeometric functions \cite{Abramowitz:1972} as a matching condition for the two asymptotic behaviors of the radial solution. In fact, the VBK approach was originally developed for the radial solution given in terms of the Heun functions \cite{Ronveaux:1995}, as well as the analytical expression for the QNMs of (3+1)-dimensional Unruh's acoustic black hole was obtained from the polynomial condition of Gauss's hypergeometric functions by Saavedra \cite{ModPhysLettA.21.1601}. Therefore, we will show that the VBK approach can also be applied for the radial solution given in terms of the hypergeometric functions, and then the radial eigenfunctions related to the QBSs has the (correct) desired behavior. We also investigate the QBSs in the (2+1)-dimensional Unruh's acoustic black hole, given the lower-dimensional properties of tabletop experiments.

This paper is organized as follows. In Sec.~\ref{4DUABH}, we introduce the metric corresponding to the four-dimensional Unruh acoustic black hole (4DUABH), solve the wave equation, and obtain the spectrum of QBSs. In Sec.~\ref{3DUABH}, we perform these same analyses in the three-dimensional Unruh acoustic black hole (3DUABH). Finally, in Sec.~\ref{Conclusions}, the conclusions are given. Here, we adopt the natural units where $G \equiv c \equiv \hbar \equiv 1$.

%
%%%%%%%%%%%%%%%%%%%%%%%%%%%%%%%%%%%%%%%%%%%%%%%%%%%%%%%%%%%%%%%%% 4DUABH
%
\section{Four-dimensional Unruh's acoustic black hole}\label{4DUABH}

In this section, we start by considering the general acoustic black-hole solution in Minkowski spacetime obtained by Unruh \cite{PhysRevLett.46.1351} (see also Refs.~\cite{ClassQuantumGrav.15.1767,LivingRevRelativity.14.3}), and then we discuss his choice for the flow velocity in order to obtain an acoustic metric that assumes just the form it has for a Schwarzschild metric near the event horizon.

The fundamental equations of motion for an irrotational fluid are given by
\begin{eqnarray}
\nabla \times \mathbf{v} & = & 0, \label{eq:irrotational}\\
\partial_{t}\rho+\nabla\cdot(\rho\mathbf{v}) & = & 0, \label{eq:Continuity}\\
\rho[\partial_{t}\mathbf{v}+(\mathbf{v}\cdot\nabla)\mathbf{v}] & = & -\nabla p, \label{eq:Euler}
\end{eqnarray}
where $\mathbf{v}$, $\rho$, and $p$ are the velocity, density, and pressure of the fluid, respectively. Next, we introduce the velocity potential $\Psi$, such that $\mathbf{v}=-\nabla\Psi$, and assume the fluid as barotropic, which means that $\rho=\rho(p)$. Then, by linearizing these equations of motion around some background $(\rho_{0},p_{0},\Psi_{0})$, namely,
\begin{eqnarray}
\rho & = & \rho_{0}+\epsilon\rho_{1}, \label{eq:rho}\\
p & = & p_{0}+\epsilon p_{1}, \label{eq:p}\\
\Psi & = & \Psi_{0}+\epsilon\Psi_{1}, \label{eq:psi}
\label{eq2:Madelung_representation}
\end{eqnarray}
we get the following wave equation:
\begin{equation}
-\partial_{t}\biggl[\frac{\partial \rho}{\partial p}\rho_{0}(\partial_{t}\Psi_{1}+\mathbf{v}_{0}\cdot\nabla\Psi_{1})\biggr]+\nabla\cdot\biggl[\rho_{0}\nabla\Psi_{1}-\frac{\partial \rho}{\partial p}\rho_{0}\mathbf{v}_{0}(\partial_{t}\Psi_{1}+\mathbf{v}_{0}\cdot\nabla\Psi_{1})\biggr]=0,
\label{eq:wave_equation_Visser}
\end{equation}
where the local speed of sound, $c_{s}$, is defined by
\begin{equation}
c_{s}^{-2} \equiv \frac{\partial \rho}{\partial p}.
\label{eq:sound}
\end{equation}
The wave equation (\ref{eq:wave_equation_Visser}) describes the propagation of the linearized scalar potential $\Psi_{1}$; that is, it governs the propagation of the phase fluctuations as weak excitations in a homogeneous stationary condensate, which can be rewritten as a wave equation in a curved spacetime
\begin{equation}
\frac{1}{\sqrt{-g}}\partial_{\mu}(g^{\mu\nu}\sqrt{-g}\partial_{\nu}\Psi_{1})=0.
\label{eq:Klein-Gordon_acoustic}
\end{equation}
Note that this wave equation is similar to the covariant Klein-Gordon equation with zero mass, where the acoustic line element can be written as
\begin{equation}
ds^{2} = g_{\mu\nu}dx^{\mu}dx^{\nu} = \frac{\rho_{0}}{c_{s}}[-c_{s}^{2}\ dt^{2}+(dx^{i}-v_{0}^{i}\ dt)\delta_{ij}(dx^{j}-v_{0}^{j}\ dt)].
\label{eq:acoustic_metric}
\end{equation}

Now, Unruh assumed the background flow as a spherically symmetric, stationary, and convergent fluid, so that the acoustic metric given by Eq.~(\ref{eq:acoustic_metric}) becomes
\begin{equation}
ds^{2}=\frac{\rho_{0}}{c_{s}}\biggl\{-[c_{s}^{2}-(v_{0}^{r})^{2}]dt^{2}+\frac{c_{s}}{c_{s}^{2}-(v_{0}^{r})^{2}}dr^{2}+r^{2}(d\theta^{2}+\sin^{2}\theta\ d\phi^{2})\biggr\},
\label{eq:Unruh_acoustic_metric}
\end{equation}
where the following coordinate transformation was performed:
\begin{equation}
t \rightarrow t+\int\frac{v_{0}^{r}}{c_{s}^{2}-(v_{0}^{r})^{2}}dr,
\label{eq:coordinate_transformation}
\end{equation}
with $v_{0}^{r} = v_{0}^{r}(r)$. Next, if the background flow smoothly exceeds the velocity of sound at the sonic horizon $r=r_{h}$, the radial component of the flow velocity can be expanded as
\begin{equation}
v_{0}^{r}(r)=-c_{s}+a(r-r_{h})+\mathcal{O}(r-r_{h})^{2},
\label{eq:fluid_velocity}
\end{equation}
where the control (tuning) parameter $a$ is defined as
\begin{equation}
a=(\nabla\cdot\mathbf{v})|_{r=r_{h}}.
\label{eq:a_4DUABH}
\end{equation}
Therefore, we can write the line element for a 4DUABH as
\begin{equation}
ds^{2}=\frac{\rho_{0}}{c_{s}}\biggl[-f(r)\ dt^{2}+\frac{c_{s}}{f(r)}\ dr^{2}+r^{2}(d\theta^{2}+\sin^{2}\theta\ d\phi^{2})\biggl],
\label{eq:metric_4DUABH}
\end{equation}
where the acoustic metric function (or warp factor) $f(r)$ has the following form:
\begin{equation}
f(r)=2c_{s}a(r-r_{h}),
\label{eq:metric_function_4DUABH}
\end{equation}
with $r_{h}$ being the acoustic event horizon, that is, the outermost marginally trapped surface for outgoing phonons, which is where the velocity of the fluid reaches the speed of sound.

In what follows, we will analyze the motion of massless scalar particles propagating at the external region of the 4DUABH. We adopt the VBK approach to obtain the exact analytical solution for the Klein-Gordon equation (\ref{eq:Klein-Gordon_acoustic}) and then use it to find the spectrum of QBS frequencies. From now on, for simplicity and without loss of generality, we will fix $c_{s}=1$, and assume the density of the fluid $\rho_{0}$ as a constant.

%
%%%%%%%%%%%%%%%%%%%%%%%%%%%%%%%%%%%%%%%%%%%%%%%%%%%%%%%%%%%%%%%%% Exact analytical solution to the wave equation
%
\subsection{Exact analytical solution to the wave equation}\label{WE}

We are interested in some basic characteristics of these 4DUABHs, in particular the ones related to their interaction with quantum scalar fields, including the Hawking-Unruh radiation and the QBSs. In order to perform these studies, we have to solve the wave equation Eq.~(\ref{eq:Klein-Gordon_acoustic}). To do this, due to the spherical symmetry, we can use the following separation ansatz
\begin{equation}
\Psi_{1}(t,r,\theta,\phi)=\mbox{e}^{-i \omega t}U(r)Y_{lm}(\theta,\phi),
\label{eq:ansatz_4DUABH}
\end{equation}
where $\omega$ is the frequency (energy, in the natural units), $U(r)=R(r)/r$ is the radial function, and $Y_{lm}(\theta,\phi)$ are the spherical harmonic functions, with $l$ and $m$ being the angular (or polar) and magnetic (or azimuthal) quantum numbers, respectively, such that $l=0,1,2,\ldots$ and $-l \leq m \leq l$. Thus, by substituting Eqs.~(\ref{eq:metric_4DUABH})-(\ref{eq:ansatz_4DUABH}) into Eq.~(\ref{eq:Klein-Gordon_acoustic}), we obtain the radial equation given by
\begin{equation}
\frac{d^{2}R(r)}{dr^{2}}+\biggl(\frac{1}{r-r_{h}}\biggr)\frac{dR(r)}{dr}+\biggl[\frac{r^{2}\omega^{2}-4a^{2}r(r-r_{h})-2a\lambda(r-r_{h})}{4a^{2}r^{2}(r-r_{h})^{2}}\biggr]R(r)=0,
\label{eq:radial_equation_4DUABH}
\end{equation}
where $\lambda=l(l+1)$ is the separation constant.

Now, we will apply the VBK approach to write the radial equation (\ref{eq:radial_equation_4DUABH}) as Gauss's hypergeometric equation without the assumption of specific boundary conditions. Then, we will impose the appropriate boundary conditions on this exact analytical radial solution in order to study the Hawking-Unruh radiation and QBS spectra. Here, we just show our main results and cordially invite the readers to find this method best described in Refs.~\cite{PhysRevD.105.045015,EurPhysJC.82.669,EurPhysJC.82.932}.

Let us define a new radial coordinate, $x$, as
\begin{equation}
x=1-\frac{r_{h}}{r},
\label{eq:radial_coordinate_4DUABH}
\end{equation}
such that the three original singularities $(r_{h},\infty)$ are moved to the points $(0,1)$. It is worth mentioning the fact that this is the same transformation for the independent variable used by Saavedra \cite{ModPhysLettA.21.1601} to study the QNMs of UABH, and by Kim \textit{et al.} \cite{PhysLettB.608.10} to analyze the decay rate and low-energy near-horizon dynamics of acoustic black holes. However, in the present work, we will choose different coefficients (and their signs) for the transformation of the dependent variable $R(x) \mapsto y(x)$, which leads to a radial solution that is more suitable for studying the Hawking-Unruh radiation and the QBS frequencies, as follows. Then, the final step is to define
\begin{equation}
R(x)=x^{A_{0}}(1-x)^{A_{1}}y(x),
\label{eq:dependent_variable_4DUABH}
\end{equation}
where the coefficients $A_{0}$ and $A_{1}$ are given by
\begin{eqnarray}
A_{0} & = & -\frac{i\omega}{2a},\label{eq:A0_4DUABH}\\
A_{1}	& = & \frac{\sqrt{4a^{2}-\omega^{2}}}{2a}.\label{eq:A1_4DUABH}
\end{eqnarray}
Thus, by substituting Eqs.~(\ref{eq:radial_coordinate_4DUABH})-(\ref{eq:A1_4DUABH}) into Eq.~(\ref{eq:radial_equation_4DUABH}), we get
\begin{equation}
x(1-x)\frac{d^{2}y(x)}{dx^{2}}+[1+2A_{0}-2(1+A_{0}+A_{1})x]\frac{dy(x)}{dx}+A_{3}y(x)=0,
\label{eq:radial_equation_2_4DUABH}
\end{equation}
where the coefficient $A_{3}$ is given by
\begin{eqnarray}
A_{3} & = & -\frac{2a^{2}r_{h}(1+A_{0}+A_{1}+2A_{0}A_{1})+a\lambda-r_{h}\omega^{2}}{2a^{2}r_{h}}.
\label{eq:A3_4DUABH}
\end{eqnarray}

The radial equation (\ref{eq:radial_equation_2_4DUABH}) has the form of Gauss's hypergeometric equation, which is given by
\begin{equation}
x(1-x)\frac{d^{2}y(x)}{dx^{2}}+[\gamma-(\alpha+\beta+1)x]\frac{dy(x)}{dx}- \alpha\beta y(x)=0,
\label{eq:hypergeometric_equation}
\end{equation}
where $y(x)={}_{2}F_{1}(\alpha,\beta;\gamma;x)$ are Gauss's hypergeometric functions \cite{Gauss:1812}. The hypergeometric series is convergent if $\gamma$ is not a negative integer (i) for all of $|x|<1$ and (ii) on the unit circle $|x|=1$ if $\mbox{Re}(\gamma-\alpha-\beta)>0$, and given by
\begin{equation}
{}_{2}F_{1}(\alpha,\beta;\gamma;x)=1+\frac{\alpha\beta}{\gamma}\frac{x}{1!}+\frac{\alpha(\alpha+1)\beta(\beta+1)}{\gamma(\gamma+1)}\frac{x^{2}}{2!}+\cdots=\sum_{n=0}^{\infty}\frac{(\alpha)_{n}(\beta)_{n}}{(\gamma)_{n}}\frac{x^{n}}{n!},
\label{eq:expansion_hypergeometric}
\end{equation}
where $(\alpha)_{n}=(\alpha+n-1)!/(\alpha-1)!$ is the Pochhammer symbol.

Therefore, the general exact solution for the radial part of the covariant massless Klein-Gordon equation in the 4DUABH spacetime can be written as
\begin{eqnarray}
U_{j}(x)=\biggl(\frac{1-x}{r_{h}}\biggr)R_{j}(x)=\biggl(\frac{1-x}{r_{h}}\biggr)x^{A_{0}}(1-x)^{A_{1}}[C_{1,j}\ y_{1,j}(x) + C_{2,j}\ y_{2,j}(x)],
\label{eq:radial_solution_4DUABH}
\end{eqnarray}
where $C_{1,j}$ and $C_{2,j}$ are constants to be determined, and $j=\{0,1\}$ labels the solution at each singular point, which are given as follows. The pair of linearly independent solutions at $x=0$ ($r=r_{h}$) is given by
\begin{eqnarray}
y_{1,0} & = & {}_{2}F_{1}(\alpha,\beta;\gamma;x),\label{eq:y10}\\
y_{2,0} & = & x^{1-\gamma}{}_{2}F_{1}(\alpha+1-\gamma,\beta+1-\gamma;2-\gamma;x),\label{eq:y20}
\end{eqnarray}
The pair of linearly independent solutions at $x=1$ ($r=\infty$) is given by
\begin{eqnarray}
y_{1,1} & = & \frac{\Gamma(\gamma)\Gamma(\gamma-\alpha-\beta)}{\Gamma(\gamma-\alpha)\Gamma(\gamma-\beta)}{}_{2}F_{1}(\alpha,\beta;\alpha+\beta-\gamma+1;1-x),\label{eq:y11}\\
y_{2,1} & = & (1-x)^{\gamma-\alpha-\beta}\frac{\Gamma(\gamma)\Gamma(\alpha+\beta-\gamma)}{\Gamma(\alpha)\Gamma(\beta)}{}_{2}F_{1}(\gamma-\alpha,\gamma-\beta;\gamma-\alpha-\beta+1;1-x).\label{eq:y21}
\end{eqnarray}
In these solutions, the parameters $\alpha$, $\beta$, and $\gamma$ are given by
\begin{eqnarray}
\alpha	& = & \frac{1}{2a}\biggl[a\biggl(1-\sqrt{1-\frac{2\lambda}{ar_{h}}}\biggr)-i\omega+\sqrt{4a^{2}-\omega^{2}}\biggr],\label{eq:alpha_4DUABH}\\
\beta		& = & \frac{1}{2a}\biggl[a\biggl(1+\sqrt{1-\frac{2\lambda}{ar_{h}}}\biggr)-i\omega+\sqrt{4a^{2}-\omega^{2}}\biggr],\label{eq:beta_4DUABH}\\
\gamma	& = & 1-\frac{i\omega}{a}.\label{eq:gamma_4DUABH}
\end{eqnarray}

Now, we will analyze the asymptotic behavior of the aforementioned general exact solution near the acoustic event horizon to investigate the Hawking-Unruh radiation. Next, we will impose some specific boundary conditions on the asymptotic behavior of the aforementioned general exact solution to compute the QBS spectra.

%
%%%%%%%%%%%%%%%%%%%%%%%%%%%%%%%%%%%%%%%%%%%%%%%%%%%%%%%%%%%%%%%%% Hawking-Unruh radiation
%
\subsection{Hawking-Unruh radiation}\label{HUR}

Near the acoustic event horizon, which means the limit when $r \rightarrow r_{h}$ (or $x \rightarrow 0$), the new radial coordinate, given by Eq.~(\ref{eq:radial_coordinate_4DUABH}), can be expanded as
\begin{equation}
x \approx \frac{r-r_{h}}{r_{h}}+\mathcal{O}(r-r_{h})^{2}.
\label{eq:x_horizon}
\end{equation}
Thus, in this limit, the radial solution, given by Eq.~(\ref{eq:radial_solution_4DUABH}), has the following asymptotic behavior
\begin{equation}
\lim_{r \rightarrow r_{h}} U_{0}(r) \sim C_{1,0}\ (r-r_{h})^{A_{0}} + C_{2,0}\ (r-r_{h})^{-A_{0}},
\label{eq:1st_condition_4DUABH}
\end{equation}
where all the remaining constants are included in $C_{1,0}$ and $C_{2,0}$, which could be determined later by assuming some additional boundary conditions, as, for example, that the radial wave function should be appropriately normalized in the range between the acoustic event horizon and the spatial infinity. Then, by using Eq.~(\ref{eq:A0_4DUABH}) we can rewrite Eq.~(\ref{eq:1st_condition_4DUABH}) as
\begin{equation}
\lim_{r \rightarrow r_{h}} U_{0}(r) \sim C_{1,0}\ \Psi_{{\rm in},0} + C_{2,0}\ \Psi_{{\rm out},0},
\label{eq:1st_wave_4DUABH}
\end{equation}
where the ingoing, $\Psi_{{\rm in},0}$, and outgoing, $\Psi_{{\rm out},0}$, scalar wave solutions near the acoustic black-hole event horizon are given, respectively, by
\begin{equation}
\Psi_{{\rm in},0}(r>r_{h}) = \mbox{e}^{-i \omega t}(r-r_{h})^{-\frac{i\omega}{2\kappa_{h}}},
\label{eq:1st_wave_in_4DUABH}
\end{equation}
and
\begin{equation}
\Psi_{{\rm out},0}(r>r_{h}) = \mbox{e}^{-i \omega t}(r-r_{h})^{+\frac{i\omega}{2\kappa_{h}}}.
\label{eq:1st_wave_out_4DUABH}
\end{equation}
Here, the gravitational acceleration on the acoustic event horizon, $\kappa_{h}$, is defined as
\begin{equation}
\kappa_{h}=\frac{1}{2} \left.\frac{df(r)}{dr}\right|_{r=r_{h}} = a.
\label{eq:grav_acc_4DUABH}
\end{equation}
Finally, by using the analytic continuation described in the VBK approach, we can obtain the relative scattering probability, $\Gamma_{h}$, and the exact spectrum of Hawking-Unruh radiation, $\bar{N}_{\omega}$, which are given, respectively, by
\begin{equation}
\Gamma_{h} = \left|\frac{\Psi_{{\rm out},0}(r>r_{h})}{\Psi_{{\rm out},0}(r<r_{h})}\right|^{2}=\mbox{e}^{-\frac{2\pi\omega}{\kappa_{h}}},
\label{eq:rel_prob_4DUABH}
\end{equation}
and
\begin{equation}
\bar{N}_{\omega} = \frac{\Gamma_{1}}{1-\Gamma_{1}}=\frac{1}{\mbox{e}^{\omega/k_{B}T_{h}}-1},
\label{eq:rad_spec_4DUABH}
\end{equation}
where $k_{B}$ is the Boltzmann constant, and $T_{h}(=\kappa_{h}/2\pi k_{B})$ is the Hawking-Unruh temperature at the acoustic event horizon. Therefore, we can conclude that the resulting spectrum of Hawking-Unruh radiation, for massless scalar particles in the 4DUABH spacetime, has a thermal character and hence it is analogous to the spectrum of black body radiation.

%
%%%%%%%%%%%%%%%%%%%%%%%%%%%%%%%%%%%%%%%%%%%%%%%%%%%%%%%%%%%%%%%%% Quasibound states
%
\subsection{Quasibound states}\label{QBSs}

The QBSs are solutions of the equation of motion, in this case given by Eq.~(\ref{eq:Klein-Gordon_acoustic}), localized at the black-hole potential well, which means that one has to impose two boundary conditions on the radial solution, namely, one related to ingoing waves at the acoustic event horizon, and the other concerning the number of particles (probability density) at the spatial infinity. After that, we need to use a matching condition for these two asymptotic behaviors of the radial solution. In this sense, in order to find the spectrum of QBSs, we follow the VBK approach, which suggests imposing the polynomial condition of the hypergeometric functions as a matching procedure.

The first boundary condition related to the QBSs means to require that the radial solution should describe only ingoing waves at the acoustic event horizon. Therefore, the first boundary condition is fully satisfied when it is imposed that $C_{2,0}=0$ in Eq.~(\ref{eq:1st_wave_4DUABH}), and hence we get
\begin{equation}
\lim_{r \rightarrow r_{h}} U_{0}(r) \sim C_{1,0}\ \Psi_{{\rm in},0}.
\label{eq:1st_boundary_4DUABH}
\end{equation}

The second boundary condition related to the QBSs means to require that the radial solution must tend to zero at spatial asymptotic infinity. Then, in the limit when $r \rightarrow \infty$ (or $x \rightarrow 1$), the new radial coordinate, given by Eq.~(\ref{eq:radial_coordinate_4DUABH}), can be expanded as
\begin{equation}
x \approx \frac{r_{h}}{r}+\mathcal{O}\biggl(\frac{1}{r}\biggr)^{2}.
\label{eq:x_infinity}
\end{equation}
Thus, in this limit, the radial solution, given by Eq.~(\ref{eq:radial_solution_4DUABH}), has the following asymptotic behavior:
\begin{equation}
\lim_{r \rightarrow \infty} U_{1}(r) \sim C_{1,1}\ \frac{1}{r^{\sigma}},
\label{eq:2nd_condition_4DUABH}
\end{equation}
where all the remaining constants are included in $C_{1,1}$ and the coefficient $\sigma$ is given by
\begin{equation}
\sigma=1+\frac{\sqrt{4a^{2}-\omega^{2}}}{2a}.
\label{eq:sigma_4DUABH}
\end{equation}
Therefore, the second boundary condition is fully satisfied when the sign of the real part of $\sigma$ is positive ($\mbox{Re}[\sigma]>0$); otherwise, if $\mbox{Re}[\sigma] < 0$ the radial solution diverges at spatial infinity. The final asymptotic behavior of the radial solution at spatial infinity will be determined when we know the values of the coefficient $\sigma$, which depends on the frequencies $\omega$, and the parameter $a$; they will be obtained in what follows by using the VBK approach.

The hypergeometric functions become polynomials of degree $n$ if they satisfy the following condition
\begin{eqnarray}
\alpha=-n,
\label{eq:hypergeometric_polynomial_condition}
\end{eqnarray}
where $n(=0,1,2,\ldots)$ is now the overtone number (or the principal quantum number). Thus, as described by the VBK approach, the polynomial condition given by Eq.~(\ref{eq:hypergeometric_polynomial_condition}) is a standard matching procedure, from which we obtain the exact spectrum of QBSs given by
\begin{equation}
\omega_{nl}^{(\pm)}=\frac{-ia(2n+1)\{\lambda+2ar_{h}[n(n+1)-1]\}\pm\{\lambda+2ar_{h}[n(n+1)+1]\}\sqrt{a(2\lambda-ar_{h})/r_{h}}}{2\lambda+4ar_{h}n(n+1)},
\label{eq:omega_4DUABH}
\end{equation}
where $\lambda=l(l+1)$ and the signs $\pm$ label the two solutions of the characteristic resonant equation; note that we get a second order equation for the frequency $\omega$ when we substitute the parameter $\alpha$ given by Eq.~(\ref{eq:alpha_4DUABH}) into the polynomial condition given by Eq.~(\ref{eq:hypergeometric_polynomial_condition}).

In Table \ref{tab:I_4DUABH} we present some values of the QBS frequencies $\omega_{nl}^{(\pm)}$, and their corresponding coefficient $\sigma_{nl}^{(\pm)}$, as functions of the acoustic event horizon $r_{h}$. From Table \ref{tab:I_4DUABH}, we can conclude that both solutions are physically admissible, since the real part of the coefficient $\sigma$ is positive in these cases, and therefore they represent QBS frequencies for massless scalar fields in the 4DUABH spacetime.

\begin{figure}%[b]
	\centering
	\includegraphics[width=1\columnwidth]{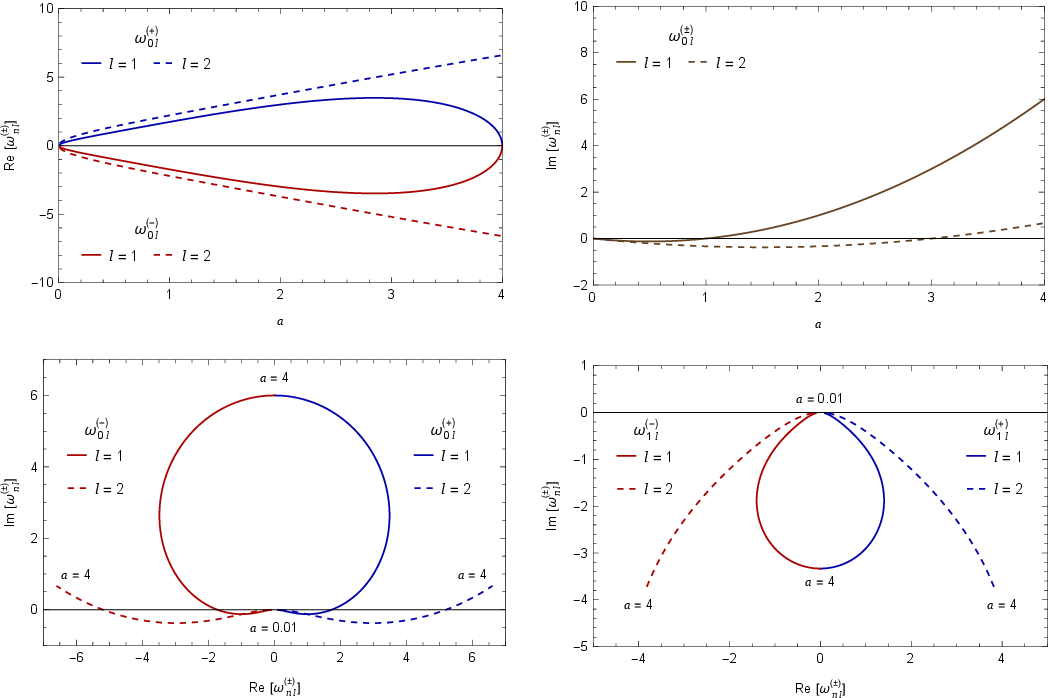}
	\caption{Top panel: Real (left) and imaginary (right) parts of the fundamental $n=0$ massless scalar QBSs of a 4DUABH with $r_{h}=1$ and varying control parameter $a$. Bottom panel: Fundamental $n=0$ (left) and first overtone $n=1$ (right) massless scalar QBS phase space of a 4DUABH with $r_{h}=1$ and varying control parameter $a$.}
	\label{fig:Fig1_QBSs_4DUABH}
\end{figure}

\begin{figure}%[t]
	\centering
	\includegraphics[width=1\columnwidth]{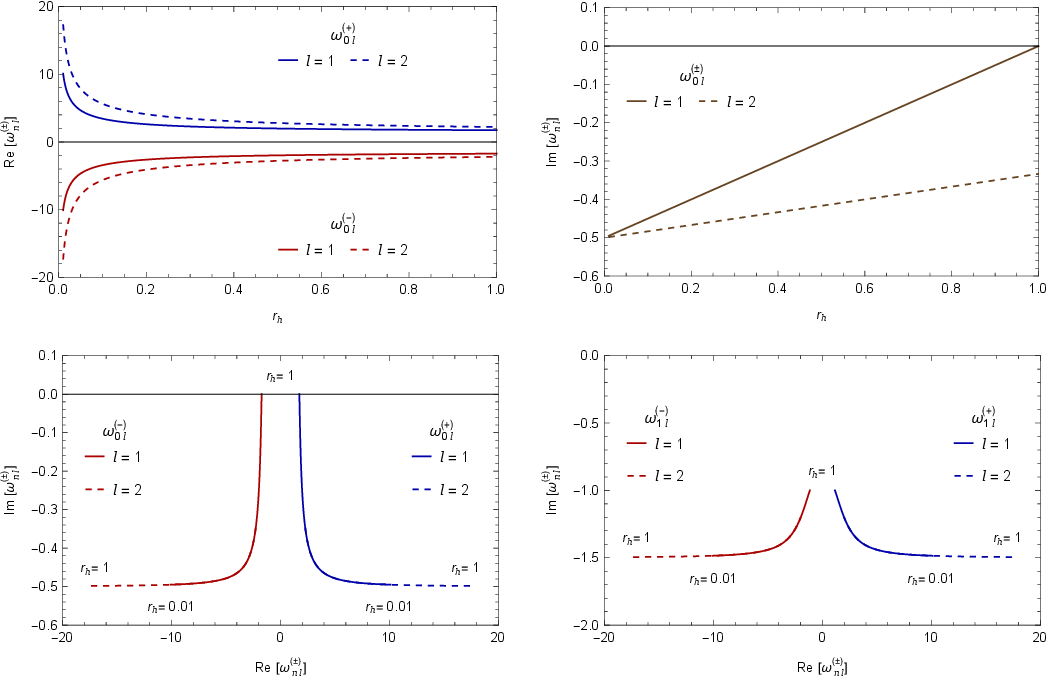}
	\caption{Top panel: Real (left) and imaginary (right) parts of the fundamental $n=0$ massless scalar QBSs of a 4DUABH with $a=1$ and varying acoustic event horizon $r_{h}$. Bottom panel: Fundamental $n=0$ (left) and first overtone $n=1$ (right) massless scalar QBS phase space of a 4DUABH with $a=1$ and varying acoustic event horizon $r_{h}$.}
	\label{fig:Fig2_QBSs_4DUABH}
\end{figure}

In Figs.~\ref{fig:Fig1_QBSs_4DUABH} and \ref{fig:Fig2_QBSs_4DUABH} we also present the behavior of these QBS frequencies $\omega_{nl}^{(\pm)}$, as functions of the control parameter $a$ and the acoustic event horizon $r_{h}$, respectively.

\begin{table}%[t]
	\caption{The massless scalar QBS frequencies $\omega_{nl}^{(\pm)}$, and their corresponding coefficient $\mbox{Re}[\sigma_{nl}^{(\pm)}]$, in the 4DUABH spacetime. We focus on the fundamental mode $n=0$, for $l=1$ and $a=1$.}
	\label{tab:I_4DUABH}
	\begin{tabular}{c|c|c|c|c}
		\hline\noalign{\smallskip}
		$r_{h}$ & $\omega_{01}^{(+)}$  & $\mbox{Re}[\sigma_{01}^{(+)}]$ & $\omega_{01}^{(-)}$ & $\mbox{Re}[\sigma_{01}^{(-)}]$ \\
		\noalign{\smallskip}\hline\noalign{\smallskip}
		0.01    & $10.08737-0.495000i$ & $1.252500$ & $-10.08737-0.495000i$ & $1.252500$ \\
		0.11    & $3.300437-0.445000i$ & $1.277500$ & $-3.300437-0.445000i$ & $1.277500$ \\
		0.21    & $2.570191-0.395000i$ & $1.302500$ & $-2.570191-0.395000i$ & $1.302500$ \\
		0.31    & $2.259819-0.345000i$ & $1.327500$ & $-2.259819-0.345000i$ & $1.327500$ \\
		0.41    & $2.086145-0.295000i$ & $1.352500$ & $-2.086145-0.295000i$ & $1.352500$ \\
		0.51    & $1.975034-0.245000i$ & $1.377500$ & $-1.975034-0.245000i$ & $1.377500$ \\
		0.61    & $1.897714-0.195000i$ & $1.402500$ & $-1.897714-0.195000i$ & $1.402500$ \\
		0.71    & $1.840496-0.145000i$ & $1.427500$ & $-1.840496-0.145000i$ & $1.427500$ \\
		0.81    & $1.795980-0.095000i$ & $1.452500$ & $-1.795980-0.095000i$ & $1.452500$ \\
		0.91    & $1.759794-0.045000i$ & $1.477500$ & $-1.759794-0.045000i$ & $1.477500$ \\
		\noalign{\smallskip}\hline
	\end{tabular}
\end{table}

From these results, we can conclude that the massless scalar QBS frequencies $\omega_{nl}^{(\pm)}$ in a 4DUABH spacetime have the following symmetry:
\begin{equation}
\omega_{nl}^{(+)}=-[\omega_{nl}^{(-)}]^{*},
\label{eq:QBSs_symmetry_4DUABH}
\end{equation}
where ``$^{*}$'' denotes complex conjugation. This symmetry indicates that these two solutions have opposite oscillation frequencies, $\mbox{Re}[\omega_{nl}^{(+)}]=-\mbox{Re}[\omega_{nl}^{(-)}]$, and the same decay rates, $\mbox{Im}[\omega_{nl}^{(+)}]=\mbox{Im}[\omega_{nl}^{(-)}]$. This kind of symmetry may describe the simultaneous particle-antiparticle creation under phonon interaction \cite{PhysLett.52B.437,LettNuovoCim.15.257}.

%
%%%%%%%%%%%%%%%%%%%%%%%%%%%%%%%%%%%%%%%%%%%%%%%%%%%%%%%%%%%%%%%%% Radial wave eigenfunctions
%
\subsection{Radial wave eigenfunctions}\label{RWE}

Now, we present the radial wave eigenfunctions related to the massless scalar QBS frequencies in the 4DUABH background. To this end, we also follow the VBK approach (for details, please see Refs.~\cite{AnnPhys.373.28,PhysRevD.104.024035}). As we explained before, these radial wave eigenfunctions are related to the polynomial condition of Gauss's hypergeometric functions.

Therefore, the QBS radial wave eigenfunctions for massless scalar fields propagating in a 4DUABH spacetime are given by
\begin{equation}
U_{nl}^{(\pm)}(x)=C_{nl}^{(\pm)}\biggl(\frac{1-x}{r_{h}}\biggr)x^{A_{0}}(1-x)^{A_{1}}{}_{2}F_{1}(-n,\beta;\gamma;x),
\label{eq:RWE_4DUABH}
\end{equation}
where $C_{nl}^{(\pm)}$ is a constant to be determined, ${}_{2}F_{1}(-n,\beta;\gamma;x)$ are Gauss's hypergeometric polynomials and the signs $(\pm)$ are related to $\omega_{nl}^{(\pm)}$.

\begin{figure}%[b]
	\centering
	\includegraphics[width=1\columnwidth]{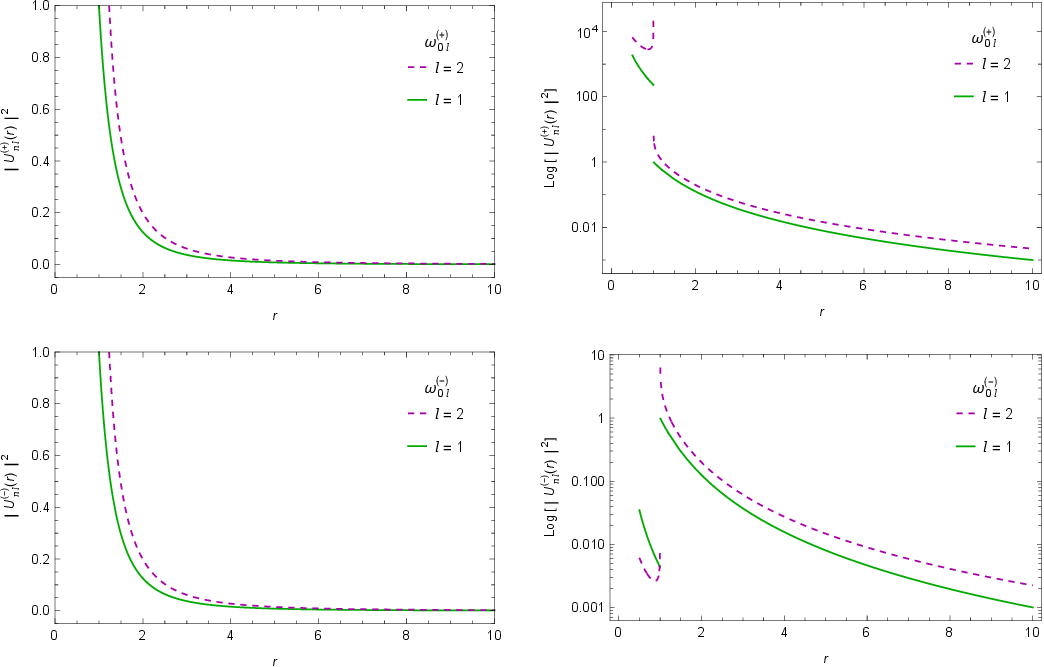}
	\caption{Top panel: The first two squared radial wave eigenfunctions (left) and their log-scale plots (right) of the fundamental $n=0$ massless scalar QBS frequencies $\omega_{nl}^{(+)}$ of a 4DUABH with $a=1$ and $r_{h}=1$, and varying radial coordinate $r$. The units are in multiples of $C_{nl}^{(+)}$.	Bottom panel: The first two squared radial wave eigenfunctions (left) and their log-scale plots (right) of the fundamental $n=0$ massless scalar QBS frequencies $\omega_{nl}^{(-)}$ of a 4DUABH with $a=1$ and $r_{h}=1$, and varying radial coordinate $r$. The units are in multiples of $C_{nl}^{(-)}$.}
	\label{fig:Fig3_Eigenfunctions_4DUABH}
\end{figure}

In Fig.~\ref{fig:Fig3_Eigenfunctions_4DUABH} we present the first two squared radial wave eigenfunctions. We observe that these radial solutions tend to zero at spatial infinity and diverge at the acoustic event horizon, which therefore represent QBSs. Note that the radial wave eigenfunctions reach a maximum value (at the acoustic event horizon $r_{h}=1$) and then cross into the acoustic black hole.

%
%%%%%%%%%%%%%%%%%%%%%%%%%%%%%%%%%%%%%%%%%%%%%%%%%%%%%%%%%%%%%%%%%%%%%%%%%%%%%%%%%%%%%%%%%%%%%% 3DUABH
%
\section{Three-dimensional Unruh's acoustic black hole}\label{3DUABH}

In this section, we discuss the Hawking-Unruh radiation and the QBSs of a 3DUABH spacetime by following the VBK approach. In the present case, we choose $\theta=\pi/2$, and then the acoustic metric given by Eq.~(\ref{eq:metric_4DUABH}) can be rewritten as
\begin{equation}
ds^{2}=\frac{\rho_{0}}{c_{s}}\biggl[-f(r)\ dt^{2}+\frac{c_{s}}{f(r)}\ dr^{2}+r^{2}\ d\phi^{2}\biggl].
\label{eq:metric_3DUABH}
\end{equation}

%
%%%%%%%%%%%%%%%%%%%%%%%%%%%%%%%%%%%%%%%%%%%%%%%%%%%%%%%%%%%%%%%%% Exact analytical solution to the wave equation
%
\subsection{Exact analytical solution to the wave equation}\label{WE_3DUABH}

In this case, we use the following separation ansatz:
\begin{equation}
\Psi_{1}(t,r,\phi)=\mbox{e}^{-i \omega t}U(r)\mbox{e}^{i m \phi},
\label{eq:ansatz_3DUABH}
\end{equation}
where $U(r)=R(r)/\sqrt{r}$ is now the radial function, and $m$ is now an integer (angular eigenvalue) such that $-\infty \leq m \leq +\infty$.

Therefore, the general exact solution for the radial part of the covariant massless Klein-Gordon equation in the 3DUABH spacetime can be written as
\begin{eqnarray}
U_{j}(x)=\sqrt{\frac{1-x}{r_{h}}}R_{j}(x)=\sqrt{\frac{1-x}{r_{h}}}x^{A_{0}}(1-x)^{A_{1}}[C_{1,j}\ y_{1,j}(x) + C_{2,j}\ y_{2,j}(x)],
\label{eq:radial_solution_3DUABH}
\end{eqnarray}
where $C_{1,j}$ and $C_{2,j}$ are constants to be determined, and $j=\{0,1\}$ labels the solution at each singular point, which are given by Eqs.~(\ref{eq:y10})-(\ref{eq:y21}). In these solutions, the coefficients $A_{0}$, and $A_{1}$, and the parameters $\alpha$, $\beta$, and $\gamma$ are now given by
\begin{eqnarray}
A_{0}   & = & -\frac{i\omega}{2a},\label{eq:A0_3DUABH}\\
A_{1}	  & = & \frac{\sqrt{a^{2}-\omega^{2}}}{2a},\label{eq:A1_3DUABH}\\
\alpha	& = & \frac{1}{2a}\biggl[a-i\omega+\sqrt{a^{2}-\omega^{2}}-\frac{i\sqrt{2}a|m|}{\sqrt{ar_{h}}}\biggr],\label{eq:alpha_3DUABH}\\
\beta		& = & \frac{1}{2a}\biggl[a-i\omega+\sqrt{a^{2}-\omega^{2}}+\frac{i\sqrt{2}a|m|}{\sqrt{ar_{h}}}\biggr],\label{eq:beta_3DUABH}\\
\gamma	& = & 1-\frac{i\omega}{a}.\label{eq:gamma_3DUABH}
\end{eqnarray}

%
%%%%%%%%%%%%%%%%%%%%%%%%%%%%%%%%%%%%%%%%%%%%%%%%%%%%%%%%%%%%%%%%% Hawking-Unruh radiation
%
\subsection{Hawking-Unruh radiation}\label{HUR_3DUABH}

In this case, the relative scattering probability, $\Gamma_{h}$, and the exact spectrum of Hawking-Unruh radiation, $\bar{N}_{\omega}$, which are given, respectively, by
\begin{equation}
\Gamma_{h} = \left|\frac{\Psi_{{\rm out},0}(r>r_{h})}{\Psi_{{\rm out},0}(r<r_{h})}\right|^{2}=\mbox{e}^{-\frac{2\pi\omega}{\kappa_{h}}},
\label{eq:rel_prob_3DUABH}
\end{equation}
and
\begin{equation}
\bar{N}_{\omega} = \frac{\Gamma_{1}}{1-\Gamma_{1}}=\frac{1}{\mbox{e}^{\omega/k_{B}T_{h}}-1},
\label{eq:rad_spec_3DUABH}
\end{equation}
which are the same results as in the case of a 4DUABH.

%
%%%%%%%%%%%%%%%%%%%%%%%%%%%%%%%%%%%%%%%%%%%%%%%%%%%%%%%%%%%%%%%%% Quasibound states
%
\subsection{Quasibound states}\label{QBSs_3DUABH}

In this case, the radial solution, given by Eq.~(\ref{eq:radial_solution_3DUABH}), has the following asymptotic behavior at spatial infinity
\begin{equation}
\lim_{r \rightarrow \infty} U_{1}(r) \sim C_{1,1}\ \frac{1}{r^{\sigma}},
\label{eq:2nd_condition_3DUABH}
\end{equation}
where all the remaining constants are included in $C_{1,1}$, and the coefficient $\sigma$ is now given by
\begin{equation}
\sigma=\frac{1}{2}+\frac{\sqrt{a^{2}-\omega^{2}}}{2a}.
\label{eq:sigma_3DUABH}
\end{equation}

Then, from the polynomial condition given by Eq.~(\ref{eq:hypergeometric_polynomial_condition}), we obtain the exact spectrum of QBSs given by
\begin{equation}
\omega_{nm}^{(\pm)}=\frac{-ia(2n+1)\sqrt{r_{h}}[m^{2}+2ar_{h}n(n+1)]\pm\sqrt{2a}\{m^{3}+mar_{h}[2n(n+1)+1]\}}{\sqrt{r_{h}}[2m^{2}+ar_{h}(2n+1)^{2}]}.
\label{eq:omega_3DUABH}
\end{equation}

In Table \ref{tab:II_3DUABH} we present some values of the QBS frequencies $\omega_{nm}^{(\pm)}$, and their corresponding coefficient $\sigma_{nm}^{(\pm)}$, as functions of the acoustic event horizon $r_{h}$. From Table \ref{tab:II_3DUABH}, we can conclude that both solutions are physically admissible, since the real part of the coefficient $\sigma$ is positive in these cases, and therefore they represent QBS frequencies for massless scalar fields in the 3DUABH spacetime.

\begin{figure}%[b]
	\centering
	\includegraphics[width=1\columnwidth]{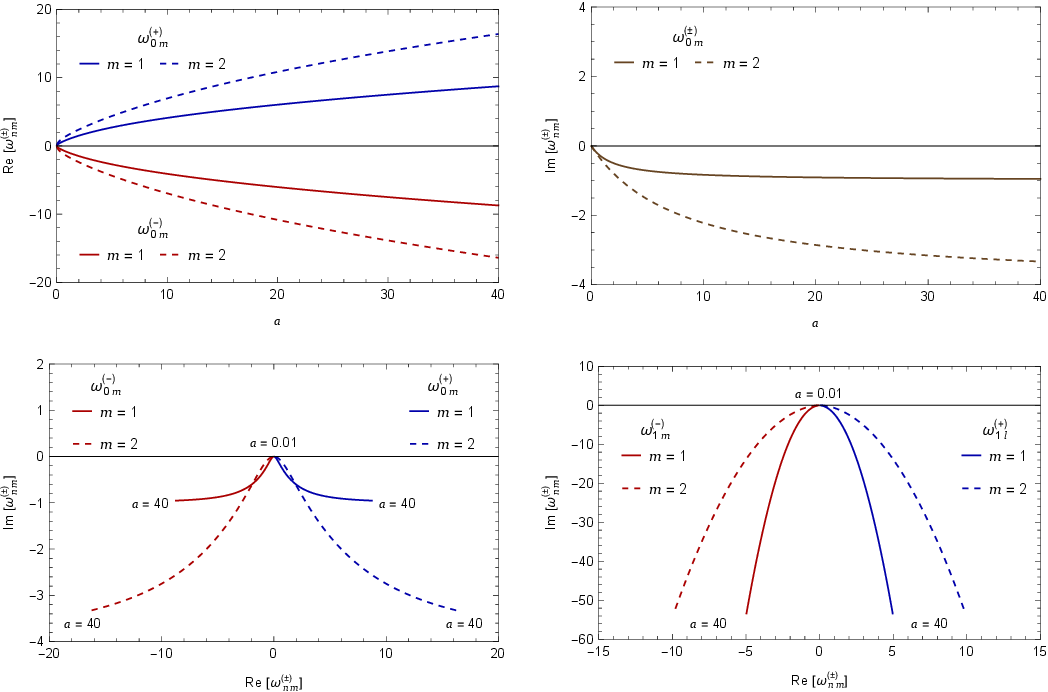}
	\caption{Top panel: Real (left) and imaginary (right) parts of the fundamental $n=0$ massless scalar QBSs of a 3DUABH with $r_{h}=1$ and varying control parameter $a$. Bottom panel: Fundamental $n=0$ (left) and first overtone $n=1$ (right) massless scalar QBS phase space of a 3DUABH with $r_{h}=1$ and varying control parameter $a$.}
	\label{fig:Fig4_QBSs_3DUABH}
\end{figure}

\begin{figure}%[t]
	\centering
	\includegraphics[width=1\columnwidth]{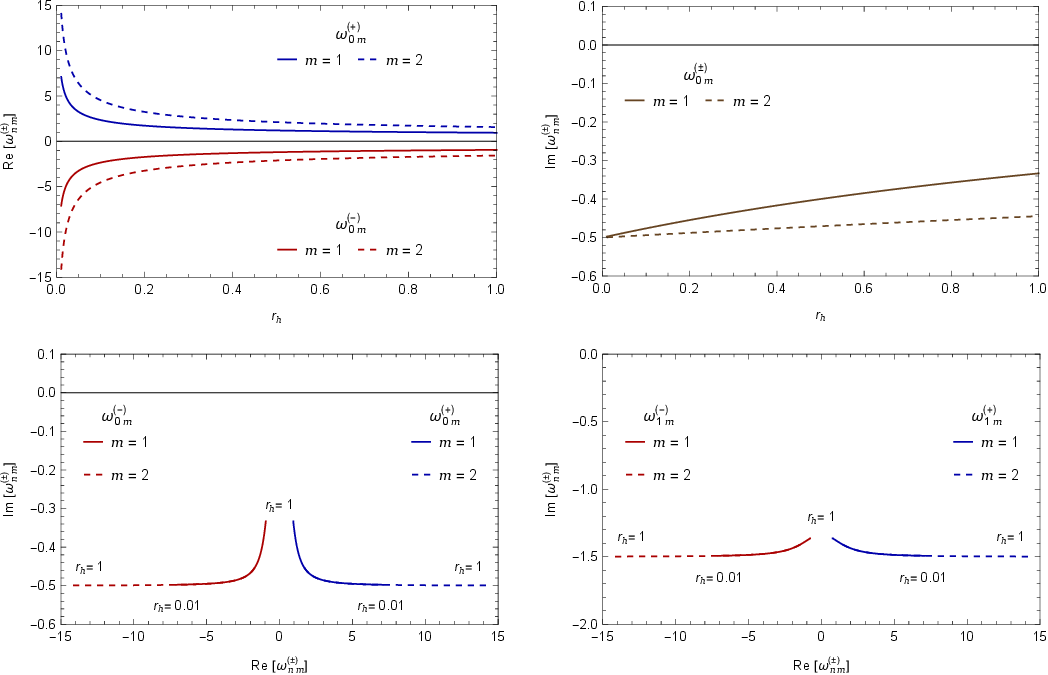}
	\caption{Top panel: Real (left) and imaginary (right) parts of the fundamental $n=0$ massless scalar QBSs of a 3DUABH with $a=1$ and varying acoustic event horizon $r_{h}$. Bottom panel: Fundamental $n=0$ (left) and first overtone $n=1$ (right) massless scalar QBS phase space of a 3DUABH with $a=1$ and varying acoustic event horizon $r_{h}$.}
	\label{fig:Fig5_QBSs_3DUABH}
\end{figure}

In Figs.~\ref{fig:Fig4_QBSs_3DUABH} and \ref{fig:Fig5_QBSs_3DUABH} we also present the behavior of these QBS frequencies $\omega_{nm}^{(\pm)}$, as functions of the control parameter $a$, and the acoustic event horizon $r_{h}$, respectively.

\begin{table}%[t]
	\caption{The massless scalar QBS frequencies $\omega_{nm}^{(\pm)}$, and their corresponding coefficient $\mbox{Re}[\sigma_{nm}^{(\pm)}]$, in the 3DUABH spacetime. We focus on the fundamental mode $n=0$, for $m=1$ and $a=1$.}
	\label{tab:II_3DUABH}
	\begin{tabular}{c|c|c|c|c}
		\hline\noalign{\smallskip}
		$r_{h}$ & $\omega_{01}^{(+)}$  & $\mbox{Re}[\sigma_{01}^{(+)}]$ & $\omega_{01}^{(-)}$ & $\mbox{Re}[\sigma_{01}^{(-)}]$ \\
		\noalign{\smallskip}\hline\noalign{\smallskip}
		0.01    & $7.106247-0.497512i$ & $0.751244$ & $-7.106247-0.497512i$ & $0.751244$ \\
		0.11    & $2.243154-0.473934i$ & $0.763033$ & $-2.243154-0.473934i$ & $0.763033$ \\
		0.21    & $1.689657-0.452489i$ & $0.773756$ & $-1.689657-0.452489i$ & $0.773756$ \\
		0.31    & $1.440434-0.432900i$ & $0.783550$ & $-1.440434-0.432900i$ & $0.783550$ \\
		0.41    & $1.292186-0.414938i$ & $0.792531$ & $-1.292186-0.414938i$ & $0.792531$ \\
		0.51    & $1.191333-0.398406i$ & $0.800797$ & $-1.191333-0.398406i$ & $0.800797$ \\
		0.61    & $1.116954-0.383142i$ & $0.808429$ & $-1.116954-0.383142i$ & $0.808429$ \\
		0.71    & $1.059041-0.369004i$ & $0.815498$ & $-1.059041-0.369004i$ & $0.815498$ \\
		0.81    & $1.012150-0.355872i$ & $0.822064$ & $-1.012150-0.355872i$ & $0.822064$ \\
		0.91    & $0.973049-0.343643i$ & $0.828179$ & $-0.973049-0.343643i$ & $0.828179$\\
		\noalign{\smallskip}\hline
	\end{tabular}
\end{table}

From these results, we can also conclude that the massless scalar QBS frequencies $\omega_{nm}^{(\pm)}$ in a 3DUABH spacetime have the symmetry given by Eq.~(\ref{eq:QBSs_symmetry_4DUABH}).

%
%%%%%%%%%%%%%%%%%%%%%%%%%%%%%%%%%%%%%%%%%%%%%%%%%%%%%%%%%%%%%%%%% Radial wave eigenfunctions
%
\subsection{Radial wave eigenfunctions}\label{RWE_3DUABH}

In this case, the QBS radial wave eigenfunctions for massless scalar fields propagating in a 3DUABH spacetime are given by
\begin{equation}
U_{nm}^{(\pm)}(x)=C_{nm}^{(\pm)}\sqrt{\frac{1-x}{r_{h}}}x^{A_{0}}(1-x)^{A_{1}}{}_{2}F_{1}(-n,\beta;\gamma;x),
\label{eq:RWE_3DUABH}
\end{equation}
where $C_{nm}^{(\pm)}$ is a constant to be determined, and the signs $(\pm)$ are related to $\omega_{nm}^{(\pm)}$.

\begin{figure}%[b]
	\centering
	\includegraphics[width=1\columnwidth]{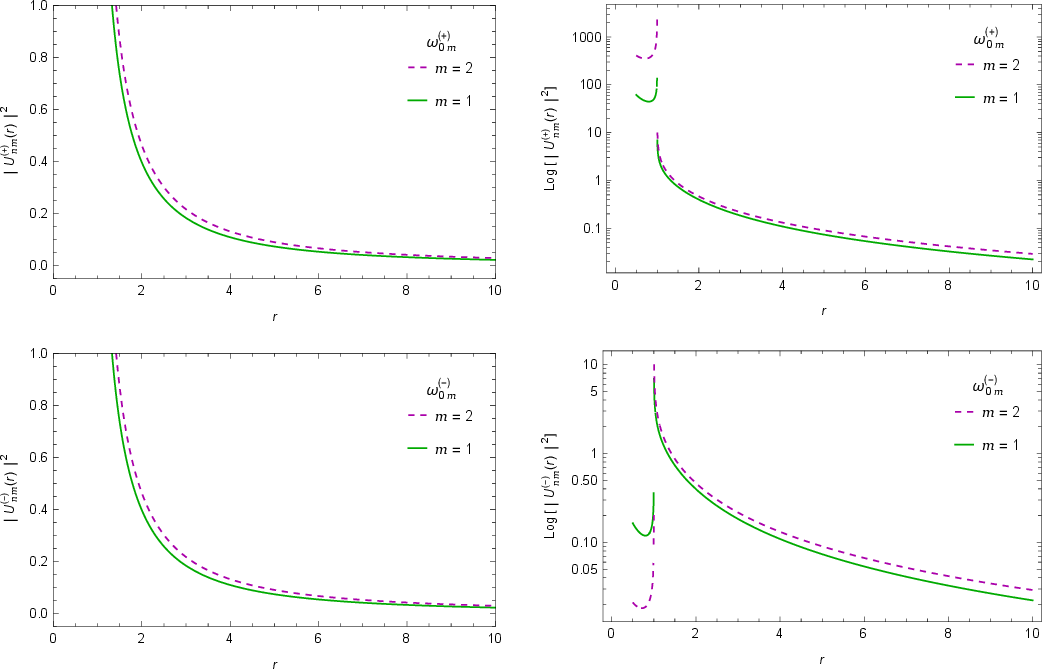}
	\caption{Top panel: The first two squared radial wave eigenfunctions (left) and their log-scale plots (right) of the fundamental $n=0$ massless scalar QBS frequencies $\omega_{nm}^{(+)}$ of a 3DUABH with $a=1$ and $r_{h}=1$, and varying radial coordinate $r$. The units are in multiples of $C_{nm}^{(+)}$.	Bottom panel: The first two squared radial wave eigenfunctions (left) and their log-scale plots (right) of the fundamental $n=0$ massless scalar QBS frequencies $\omega_{nm}^{(-)}$ of a 3DUABH with $a=1$ and $r_{h}=1$, and varying radial coordinate $r$. The units are in multiples of $C_{nm}^{(-)}$.}
	\label{fig:Fig6_Eigenfunctions_3DUABH}
\end{figure}

In Fig.~\ref{fig:Fig6_Eigenfunctions_3DUABH} we present the first two squared radial wave eigenfunctions. We observe that these radial solutions tend to zero at spatial infinity and diverge at the acoustic event horizon, which therefore represent QBSs in a 3DUABH spacetime.

%
%%%%%%%%%%%%%%%%%%%%%%%%%%%%%%%%%%%%%%%%%%%%%%%%%%%%%%%%%%%%%%%%%%%%%%%%%%%%%%%%%%%%%%%%%%%%%% Conclusions
%
\section{Conclusions}\label{Conclusions}

In this work we obtained exact analytical solutions for the covariant massless Klein-Gordon equation in both four-dimensional and three-dimensional Unruh's acoustic black-hole spacetimes. The radial solutions are given in terms of Gauss's hypergeometric functions.

The study of the asymptotic behaviors of the radial solution led to the Hawking-Unruh radiation and quasibound state phenomena. Near the acoustic event horizon, the radial solution diverges by reaching a maximum value, which indicates that the massless scalar particles cross into the acoustic black hole. On the other hand, far from the acoustic black hole at the asymptotic spatial infinity, the radial solution tends to zero; that is, the probability of finding any particles there is null.

We obtained the spectrum of quasibound state frequencies for massless scalar particles propagating in both four-dimensional and three-dimensional Unruh's acoustic black-hole spacetimes. This becomes possible by using the VBK approach, which was developed to study the quasibound states. Therefore, these results are new and, to our knowledge, there are no similar results in the literature for the backgrounds under consideration.

We obtained a set of eigenfrequencies that are solutions of a characteristic resonant equation, which were derived from the polynomial condition of Gauss's hypergeometric functions. All of these solutions are physically acceptable. From these spectra, we also obtained the radial wave eigenfunctions.

As a future perspective, it is possible to extend our results in order to obtain a new acoustic curved Unruh's acoustic black hole embedded in Schwarzschild spacetime by using the the Gross-Pitaevskii theory \cite{NuovoCimento.20.454,SovPhysJETP.13.451}. Regarding observational prospects, we can mention that the 4DUABH embedded in a curved spacetime could describe a general black-hole background immersed in intergalactic and/or cosmological media supporting the propagation of sound waves. Concerning the 3DUABH, this kind of analog model of gravity seems to be more suitable for testing in a ground-based laboratory.

%
%%%%%%%%%%%%%%%%%%%%%%%%%%%%%%%%%%%%%%%%%%%%%%%%%%%%%%%%%%%%%%%%%%%%%%%%%%%%%%%%%%%%%%%%%%%%%% Author contributions
%
%\section*{Author contributions}
%
%H. S. Vieira: Conceptualization, investigation, methodology, software, writing.
%
%
%%%%%%%%%%%%%%%%%%%%%%%%%%%%%%%%%%%%%%%%%%%%%%%%%%%%%%%%%%%%%%%%%%%%%%%%%%%%%%%%%%%%%%%%%%%%%% Data availability
%
\section*{Data availability}

The data that support the findings of this study are available from the corresponding author upon reasonable request.

%
%%%%%%%%%%%%%%%%%%%%%%%%%%%%%%%%%%%%%%%%%%%%%%%%%%%%%%%%%%%%%%%%%%%%%%%%%%%%%%%%%%%%%%%%%%%%%% Acknowledgments
%
\begin{acknowledgments}

H.S.V. is funded by the Alexander von Humboldt-Stiftung/Foundation (Grant No. 1209836). This study was financed in part by the Conselho Nacional de Desenvolvimento Cient\'{i}fico e Tecnol\'{o}gico -- Brasil (CNPq) -- Research Project No. 150410/2022-0. The author thanks Professor K. D. Kokkotas and Dr. Kyriakos Destounis for useful discussions on topics covered in this work. It is a great pleasure to thank the Theoretical Astrophysics at T\"{u}bingen (TAT Group) for its wonderful hospitality and technological support (software licenses).

\end{acknowledgments}
%
%%%%%%%%%%%%%%%%%%%%%%%%%%%%%%%%%%%%%%%%%%%%%%%%%%%%%%%%%%%%%%%%%%%%%%%%%%%%%%%%%%%%%%%%%%%%%% thebibliography
%

%
%%%%%%%%%%%%%%%%%%%%%%%%%%%%%%%%%%%%%%%%%%%%%%%%%%%%%%%%%%%%%%%%%%%%%%%%%%%%%%%%%%%%%%%%%%%%%%
%
\end{document}